\documentclass[journal]{IEEEtran}
\usepackage{cite}
\usepackage{amsmath,amssymb,amsfonts}
\usepackage[hyphens]{url}
\usepackage{algorithm,algpseudocode}
\usepackage{mathtools}
\usepackage{graphicx}
\usepackage{textcomp}
\usepackage{xcolor}

\usepackage{flushend}
\usepackage{stfloats}
\usepackage{lipsum}
\usepackage[english]{babel}
\usepackage{csquotes}
\usepackage{interval}
\usepackage{hyperref}
\usepackage{autobreak}
\usepackage{bm}
\usepackage[caption=false,font=normalsize,labelfont=sf,textfont=sf]{subfig}
\usepackage{listofitems}
\usepackage{multirow}
\usepackage{tabularx,booktabs}

\usepackage{caption}
\usepackage{titlesec}
\usepackage{utfsym}
\usepackage{pxpgfmark}   
\newcolumntype{C}{>{\centering\arraybackslash}X} 
\titleformat{\subsubsection}[runin]
  {\normalfont\bfseries} 
  {}{0em}{}[: ]  
\titlespacing*{\subsubsection}{0pt}{0.5em}{0.5em} 

\begin{document}
\bstctlcite{IEEEexample:BSTcontrol}
\intervalconfig {
  soft open fences,
}

\title{Robust Unsupervised Network Intrusion Detection via Federated Learning with Selective Aggregation under Anomalous Sample Contamination
  \thanks{This work is supported in part by JST, PRESTO Grant Number JPMJPR2035 and JST-ALCA-Next Japan Grant Number JPMJAN24F1.}
  \thanks{Shohei Kamiguchi and Takayuki Nishio are with the School of Engineering, Institute of Science Tokyo, Tokyo 152-8550, Japan (e-mail: nishio@ict.e.titech.ac.jp).}

}

\author{
  Shohei Kamiguchi and Takayuki Nishio,~\IEEEmembership{Senior~Member,~IEEE}  
}

% {Shell \MakeLowercase{\textit{et al.}}: Article Using IEEEtran.cls for IEEE Journals}

\maketitle

\begin{abstract}
  Network intrusion detection systems (NIDS) have become essential for Internet of Things (IoT) environments, as malware targeting IoT devices continues to evolve in sophistication. Unsupervised learning approaches offer a promising direction by removing the dependency on labeled datasets. However, the common assumption that training data are entirely clean is often violated in practice, particularly when data samples are collected directly from deployed network devices, where anomalies are likely to be present in the training datasets. Such contamination degrades detection performance and highlights the need for robust unsupervised NIDS methods capable of operating effectively under contaminated unlabeled training data.
  To address this issue, we propose a robust training methodology for anomaly detection (AD) that remains effective even in the presence of unlabeled anomalies. Our method consists of two primary components. First, we exploit a known limitation of federated learning (FL)—its tendency to underrepresent minority data. By leveraging this characteristic, we attenuate the influence of anomalous data originating from a small number of compromised clients. Second, we introduce a selective aggregation mechanism during model aggregation, which quantifies the “distance” between local client models and a global reference. Specifically, we employ the Expectation-Maximization (EM) algorithm to detect and exclude client groups whose model updates significantly diverge from the majority. This selective aggregation ensures that anomalous updates do not compromise the global model.
  Experiments conducted on multiple NIDS datasets demonstrate that our method outperforms existing approaches in environments contaminated with anomalous data. Furthermore, the proposed method maintains its detection performance even as the proportion of anomalies increases.
\end{abstract}


\section{Introduction}
 
\begin{figure}[t]
  \centering

  \includegraphics[width=\columnwidth]{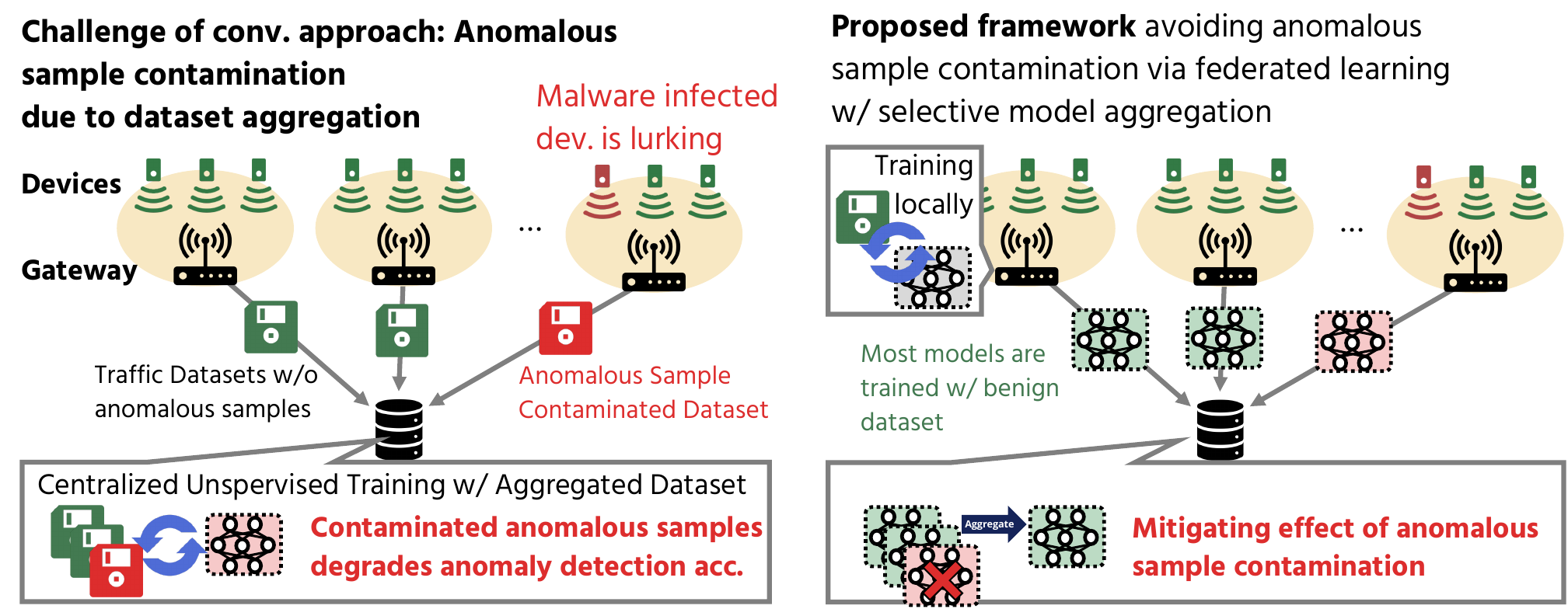}
  \caption{Challenge of conventional unsupervised NIDS, anomalous sample contamination, and proposed solution. Red-colored icons represent devices, datasets, and models affected by malware's anomalous samples, which causes performance degradation.}
  \label{fig:challenge}
\end{figure}

The widespread adoption of IoT is transforming modern digital infrastructure, with billions of devices embedded in homes, industries, and public services. While this connectivity enables significant advances in automation and data-driven functionality, it also introduces substantial security risks~\cite{HADDADPAJOUH2021100129}. Many IoT devices operate with limited computational capacity and are deployed in heterogeneous environments, making them attractive targets for malware. Attacks such as the Mirai botnet~\cite{7971869} have shown how vulnerable IoT devices can be leveraged to launch large-scale disruptions. Protocol-level studies~\cite{8688434, 8386824} further reveal how complex device interactions can be exploited. According to Nokia's Threat Intelligence Report 2023~\cite{NOKIA}, approximately $0.1\%$ of connected devices were infected monthly, with IoT devices accounting for $9\%$ of these cases. In residential networks, the infection rate has remained at $2\%$ per month since 2019. These trends highlight the urgent need for effective security mechanisms. Among them, NIDS play a critical role in detecting malicious activity within IoT traffic, offering a last line of defense in environments where endpoint protection is often insufficient.

Recent developments in NIDS have increasingly turned to machine learning (ML) techniques, which enable the automated identification of suspicious patterns in network traffic~\cite{HADDADPAJOUH2021100129}. ML-based NIDS are particularly appealing in IoT contexts, where network behavior is heterogeneous and attack vectors are often unknown. Supervised ML approaches have demonstrated strong detection performance~\cite{9424138, liEfficientFederatedLearning2023, 10.1007/978-3-031-34776-4_7}; however, their effectiveness hinges on the availability of large volumes of labeled training data. In practice, such datasets are difficult and costly to obtain, especially for newly emerging malware, for which labeled instances are often unavailable. Consequently, supervised methods face limitations in adapting to novel and evolving threats. As a more practical and scalable alternative, unsupervised NIDS have attracted growing attention~\cite{binbusayyisUnsupervisedDeepLearning2021, alomNetworkIntrusionDetection2017, verkerkenUnsupervisedMachineLearning2020}. These methods learn patterns of normal behavior and detect anomalies without the need for labeled data, making them particularly suitable for real-world IoT environments where threat landscapes are continuously shifting and labeled data remain limited.

In typical unsupervised NIDS settings, training data are assumed to consist exclusively of benign traffic. However, this assumption often breaks down in practical IoT deployments. In many cases, traffic is collected from routers or gateways in operational environments, where already-infected devices may be actively communicating on the network. This means that anomalous traffic can be inadvertently included in the training data even before any detection model is applied. Such contamination poses a significant challenge: unsupervised models trained on mixed data are likely to internalize malicious behavior as normal, leading to degraded detection performance.

Although the issue of anomalous contamination has been actively studied in the broader field of unsupervised AD~\cite{yoon2021self, 4781136}, existing methods are not directly tailored to NIDS. For instance, Qiu \textit{et al.}\cite{pmlr-v162-qiu22b} proposed a state-of-the-art (SoTA) AD approach that incorporates both inlier and outlier information into the loss function. While their method is generalizable and performs well across a range of tasks, it does not explicitly leverage the structural characteristics of IoT-based NIDS scenarios—particularly the fact that malicious activity is typically confined to a small subset of devices, while the majority of the network traffic originates from benign sources. By exploiting this property, it becomes possible to achieve more effective and targeted mitigation of contamination in unsupervised NIDS.

Building on this insight, we propose a novel unsupervised NIDS framework, FLANDRE (\textit{Federated Learning-assisted Anomalous Node Detection \& REmoval}) that explicitly incorporates this device-level structure to enhance robustness in the presence of contaminated training data. Our approach assumes a realistic IoT setting where traffic is distributed across multiple clients—each representing a group of local devices—and only a minority of clients are compromised. We employ FL to enable collaborative model training without centralizing sensitive traffic data. Crucially, we turn a well-known limitation of FL—its tendency to underrepresent minority patterns—into an advantage, naturally suppressing the influence of anomalous samples from compromised clients. Furthermore, we introduce a selective aggregation mechanism that evaluates the divergence between local and global models and excludes clients exhibiting suspicious behavior. In contrast to prior work that focuses on sample-level filtering~\cite{pmlr-v162-qiu22b}, our method performs client-level filtering throughout training, requires no labeled data, and offers improved robustness and adaptability in practical IoT-based NIDS scenarios, as illustrated in Figure~\ref{fig:challenge}.

The main contributions of this paper are summarized as follows:

\noindent
-- \textbf{Contamination-robust unsupervised NIDS framework:}
We propose a novel training framework for unsupervised NIDS that addresses anomalous contamination in training data, a practical issue in IoT environments. The framework uses FL as an intrinsic client-level anomaly suppression mechanism. Rather than employing FL solely for privacy or scalability, we exploit its inherent tendency to underrepresent minority patterns to mitigate the impact of compromised devices, under the realistic assumption that most IoT devices generate benign traffic. Furthermore, we introduce a selective aggregation mechanism that identifies and excludes suspicious clients based on model divergence, enabling robust model aggregation without requiring labeled data.

\noindent
-- \textbf{Comprehensive evaluation and comparison with SoTA:}
We conduct extensive experiments on three representative NIDS benchmark datasets, namely ToN\_IoT, CSE-CIC-IDS2018 (IDS2018), and NF-UQ-NIDS-v2 (NF-UQ-NIDS), under varying contamination settings. Our proposed method consistently outperforms both conventional and SoTA approaches, including the recent soft Latent Outlier Exposure (LOE-S) method.
In particular, under a representative contamination setting, FLANDRE achieves F1 scores of $0.969$, $0.838$, and $0.823$ on ToN\_IoT, NF-UQ-NIDS, and IDS2018, respectively, whereas LOE-S attains $0.916$, $0.742$, and $0.586$. Furthermore, FLANDRE remains consistently close to the ideal contamination-free reference, with a F1 score gap of only $0.004$ on ToN\_IoT, highlighting its robustness against anomalous contamination.

This work builds upon our previous conference publication~\cite{10000633}, which presented an initial proof-of-concept applying FL framework to solve the contamination problem in unsupervised NIDS. The prior study did not incorporate any selective aggregation strategy and included only limited evaluation. In contrast, the present paper introduces a new client removal technique, expands the scope of analysis, and provides a detailed comparison with recent SoTA methods. Code for reproducing our results is available at \url{https://github.com/nishio-laboratory/FLANDRE}.

Note that our problem setting differs from that of conventional studies on AD with FL.
Prior work has applied FL to AD~\cite{10076669}, primarily to protect training data by exchanging only model parameters. In existing FL-based AD approaches, client filtering methods are introduced to mitigate performance degradation caused by variations in data quality and distribution across clients, as well as to reduce computational and communication costs during training~\cite{liEfficientFederatedLearning2023,10.1007/978-3-031-34776-4_7}. 
In contrast, our work addresses the problem of anomalous data contamination in NIDS. We propose a training framework for constructing a robust NIDS by leveraging the inherent robustness of FL against a minority of anomalous clients, together with a client filtering algorithm designed based on this property.

\section{Related Works}\label{sec:rw}

We summarize related work in three categories: (i) unsupervised NIDS approaches and their vulnerability to training data contamination, (ii) contamination-robust unsupervised AD methods, and (iii) applications of FL in the context of AD and NIDS. Table~\ref{tab:method-comparison} provides a high-level comparison, and the following subsections discuss each category in detail.

\subsection{Unsupervised NIDS with Anomalous Data Contamination}\label{subsec:unsupervisedNIDS}

Unsupervised learning has been widely adopted in NIDS, where models are typically trained only on normal traffic to detect deviations. Binbusayyis and Vaiyapuri~\cite{binbusayyisUnsupervisedDeepLearning2021} proposed a framework combining a one-dimensional convolutional autoencoder with a one-class support vector machine (SVM). Alom and Taha~\cite{alomNetworkIntrusionDetection2017} employed restricted Boltzmann machines with K-means clustering. Verkerken~\cite{verkerkenUnsupervisedMachineLearning2020} compared several unsupervised approaches, including one-class SVM, isolation forest, principal component analysis (PCA), and autoencoders, reporting that autoencoders yielded the best results.

However, these methods generally assume that the training data are entirely clean, and overlook the risk of anomalous sample contamination—a realistic concern in IoT environments where traffic is often collected from deployed routers or gateways. Zhang \textit{et al.}\cite{zhangAnomalyBasedNetwork2006} explicitly investigated the impact of training data contamination, showing that the presence of anomalies degrades detection performance. Choi \textit{et al.}\cite{choiUnsupervisedLearningApproach2019} further analyzed the sensitivity of autoencoders to contaminated training data, though their work did not propose methods to mitigate its effects.

\subsection{Contamination-Robust Unsupervised AD}\label{subsec:unsupervisedADwithContamination}

In the broader field of unsupervised AD, several studies have proposed methods to address training data contamination. Isolation Forest~\cite{4781136} detects anomalies by measuring how easily data points can be isolated in random partitions. Yoon \textit{et al.}\cite{yoon2021self} proposed a data refinement strategy that uses ensembles of one-class classifiers to remove outliers prior to model training. Other studies have incorporated similar strategies into latent space models, such as latent support vector data description (SVDD)~\cite{gornitz2014learning} and autoencoder-based detectors~\cite{xia2015learning, beggel2020robust}.

Zhou and Paffenroth~\cite{zhou2017anomaly} introduced a robust autoencoder that learns to reconstruct only normal samples while penalizing abnormal ones, though their approach requires training an auxiliary anomaly detector. Qiu \textit{et al.}\cite{pmlr-v162-qiu22b} proposed a unified training framework that separates the treatment of inliers and outliers through tailored loss functions, achieving SoTA performance across diverse AD tasks. Their method consistently outperforms earlier approaches such as refinement-based filtering\cite{yoon2021self}, robust autoencoders~\cite{zhou2017anomaly}, and SVDD-based methods~\cite{gornitz2014learning, xia2015learning}, particularly on tabular datasets. For this reason, we consider it the current SoTA in contamination-robust unsupervised AD and include it as a baseline in our evaluation.

However, these methods has not been specifically designed and evaluated in the context of NIDS. In particular, they do not exploit the structural property often observed in NIDS scenarios—namely, that only a small subset of clients (or devices) are compromised, while the majority generate benign traffic. This motivates the need for approaches that are tailored to such structural characteristics of IoT-based NIDS environments.

\subsection{FL for AD and NIDS}\label{subsec:FLinAD}

FL has been explored in various AD settings due to its ability to enable distributed training while preserving data locality. In the context of NIDS, existing FL-based studies have focused primarily on supervised learning. Mothukur \textit{et al.}\cite{9424138} proposed a federated gated recurrent unit (GRU) model for IoT intrusion detection, where model weights—not raw data—are shared with a central server. Dynamic weighted aggregation federated learning (DAFL)~\cite{liEfficientFederatedLearning2023} incorporates a client selection mechanism based on performance to improve aggregation quality. FedGroup~\cite{10.1007/978-3-031-34776-4_7} clusters clients by traffic similarity to enable group-wise training. While promising, these approaches require labeled data and do not address the issue of contaminated training data, limiting their practicality in unsupervised NIDS scenarios.

In unsupervised FL-based AD, only a few studies exist. Nardi \textit{et al.}~\cite{10076669} proposed a method that clusters clients based on inlier similarity and performs intra-group aggregation. However, this approach assumes the ability to partition clients based on shared normal patterns, which may not be applicable in NIDS settings where the presence of compromised clients is unknown.

Unlike previous work that primarily adopts FL for privacy preservation, our study leverages FL to suppress the influence of compromised clients by exploiting its tendency to underrepresent minority data. This often-cited limitation of FL is reinterpreted here as a useful mechanism for contamination mitigation. Additionally, we introduce a selective aggregation technique based on model divergence to dynamically exclude suspicious clients from the training process, thereby enhancing the robustness of the learned model under real-world conditions.

\begin{table*}[t]
  \centering
  \renewcommand{\arraystretch}{1.2}
  \begin{tabularx}{\textwidth}{@{} l c c c c c @{}}
    \toprule
    \textbf{Method}                                                                                                             & \textbf{Strategy} & \textbf{NIDS} & \shortstack{\textbf{Contaminated}\\\textbf{Training}} 
    & \shortstack{\textbf{Unlabeled}\\\textbf{Data}} & \shortstack{\textbf{Contamination}\\\textbf{Robust}} \\
    \midrule
    \cite{binbusayyisUnsupervisedDeepLearning2021, alomNetworkIntrusionDetection2017, verkerkenUnsupervisedMachineLearning2020} &
    Unsupervised ML                                                                                                             & Yes               & No            & Yes                            & No                                                      \\

    \cite{zhangAnomalyBasedNetwork2006, choiUnsupervisedLearningApproach2019}                                                   &
    Unsupervised ML                                                                                                             & Yes               & Yes           & Yes                            & No                                                      \\

    \cite{pmlr-v162-qiu22b, yoon2021self, gornitz2014learning, xia2015learning, beggel2020robust}                               &
    Contamination-Aware ML                                                                                                      & Needs Mod.        & Yes           & Yes                            & Yes                                                     \\

    \cite{9424138, liEfficientFederatedLearning2023, 10.1007/978-3-031-34776-4_7}                                               &
    Supervised FL                                                                                                               & Yes               & No            & No                             & No                                                      \\

    \cite{10076669}                                                                                                             &
    Unsupervised FL                                                                                                             & Not for NIDS      & Yes           & Yes                            & Partial                                                 \\

    \textbf{This Work (FLANDRE)}                                                                                                &
    Unsupervised FL                                                                                                             & Yes               & Yes           & Yes                            & \textbf{Yes}                                            \\
    \bottomrule
  \end{tabularx}
  \caption{Comparison of related works by learning strategy, NIDS applicability, training data contamination, labeling needs, and robustness.}
  \label{tab:method-comparison}
\end{table*}

\section{Proposed Method: FLANDRE}\label{sec:method}
This section presents our proposed method, \textbf{FLANDRE}, which improves the robustness of unsupervised AD when training data are partially contaminated by malware traffic. The primary objective of FLANDRE is to suppress the influence of anomalous samples contributed by a small number of compromised clients—an inherent characteristic of many real-world NIDS scenarios. To achieve this, our approach leverages FL as a means to isolate and filter out such clients during training.
We begin by introducing the system model and assumptions underlying our design. We then describe the implementation of our client exclusion and aggregation mechanism, followed by theoretical analysis of its robustness under contamination.

\subsection{System Model}

We consider the problem of detecting malware activity from IoT network traffic using flow-based features such as flow duration, the amount of transferred data, and the number of packets, as provided in benchmark datasets including ToN\_IoT~\cite{toniot} and IDS2018~\cite{sharafaldinGeneratingNewIntrusion2018}.

As illustrated in Figure~\ref{fig:challenge}, IoT devices connect to gateways (GWs), which serve as observation points for monitoring network traffic and generating unlabeled datasets for unsupervised AD. NIDS are deployed at these GWs and operate using a shared AD model trained collaboratively across multiple GWs. In our framework, each GW acts as a client in an FL setting, performing a small number of local training epochs before sending model updates to a central server. The server functions as the FL parameter server, aggregating updates and isolating potentially compromised clients without requiring access to raw traffic data.

The dataset collected at the $i$-th GW is denoted by \(D_i\) and consists of flow records \(x_j \in D_i\). Each record has a ground-truth label \(y_j = 1\) for malware flows and \(y_j = 0\) for benign flows. These labels are hidden during training and are used solely for evaluation.

We assume that only a small fraction of IoT devices are infected with malware. A GW hosting at least one infected device inevitably collects anomalous flow data, contaminating its local dataset. Such GWs are referred to as \textit{compromised clients}, while those whose datasets contain only benign IoT application flows are referred to as \textit{benign clients}:
\[
  \begin{aligned}
    \text{Compromised:} & \quad \exists\, x_j \in D_i \ \text{s.t.} \ y_j = 1, \\
    \text{Benign:}      & \quad \forall\, x_j \in D_i, \ y_j = 0.
  \end{aligned}
\]
The proportion of compromised clients and the fraction of anomalous samples within each compromised client are both unknown.

In conventional centralized training, datasets from all GWs are collected at a central server to form a large training set, which can improve model performance but risks incorporating anomalous samples. In contrast, FLANDRE trains models in a distributed manner, mitigating the influence of contaminated data while preserving data locality. The training procedure is detailed in the next section.

% \begin{figure}[t]
%   \centering
%   \resizebox{\columnwidth}{!}{%
%     \input{./fig/system_model.tex}
%   }
%   \caption{System Model}
%   \label{fig:system-model}
% \end{figure}
\subsection{Overview of FLANDRE Training Process}

\begin{figure}[!t]
  \begin{algorithm}[H]
    \caption{FLANDRE}
    \label{alg:fedavg}
    \begin{algorithmic}[1]
      \Procedure{FL Server Training}{}
      \State Initialize global model $w^0$
      \For{each round $t = 0, 1, \dots, T$}
      \State Broadcasts $w^t$ to all client $i \in S$
      \For{each $i \in S$ parallelly}
      \State $\Delta w^{t+1}_i \leftarrow \text{ClientUpdate}(i, w^t)$
      \EndFor
      
    \If{$t>\mathrm{threshold}$} 
    \State $S^t \leftarrow \text{RemoveCompromisedClients}(S)$
    \Else 
    \State $S^t \leftarrow S$
    \EndIf
      \State $w^{t+1} \leftarrow w^t + \frac{1}{\lvert D_{S^t}\rvert}\sum_{i \in S^t} \lvert D_i \rvert \Delta w^{t+1}_i$
      \EndFor
      \EndProcedure
      \vspace{5pt}
      \Function{ClientUpdate}{$i, w$}

      $\hat{w} \leftarrow w - \eta \nabla L(D_i; w)$

      \State \Return $\Delta w \leftarrow  \hat{w}-w$
      \EndFunction
      \vspace{5pt}
      \Function{RemoveCompromisedClients}{$S$}

      \State $S_0,S_1 \leftarrow \mathrm{GaussianMixture}(S)$

      \If{$d(S_0) > d(S_1)$}

      \State \Return $S_1$

      \Else
      \State \Return $S_0$

      \EndIf
      \EndFunction
    \end{algorithmic}
  \end{algorithm}
\end{figure}

\begin{figure}[t]
  \centering

  \includegraphics[width=\columnwidth]{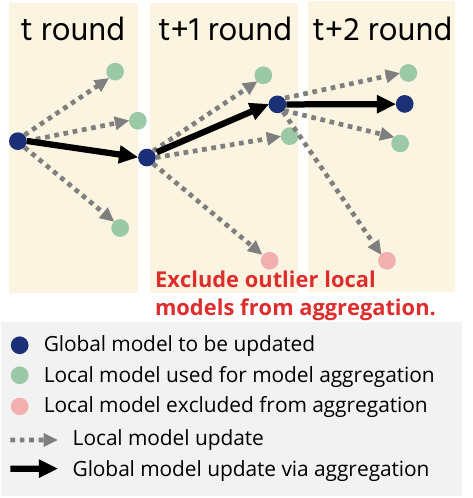}
  \caption{Illustration of the selective aggregation process in the model parameter space. Local models are divided into two clusters, one of which is identified as an outlier cluster corresponding to compromised client models and excluded from model aggregation.}
  \label{fig:fl-anomaly-moderation}
\end{figure}

To address performance degradation caused by anomalous sample contamination, our method combines FL with a selective aggregation mechanism. While FL is typically adopted to preserve client data privacy, it is also known to underrepresent minority patterns when training data are imbalanced~\cite{wang2021addressing}. In FL, such imbalance can occur not only globally across all clients but also locally within each client, making minority patterns even more prone to being overlooked compared to centralized learning.

In FLANDRE, we exploit this tendency to suppress the influence of anomalous flows, which constitute a minority class originating from a small fraction of compromised clients. Furthermore, by measuring the divergence between models trained on benign data and those affected by anomalies, and leveraging the assumption that the proportion of compromised clients is small, we exclude suspicious clients from model aggregation. This two-step process mitigates the impact of contamination while maintaining collaborative training among the remaining clients.

Algorithm~\ref{alg:fedavg} outlines the training process of the proposed method. It follows the standard Federated Averaging (FedAvg) procedure \cite{pmlr-v54-mcmahan17a}, with two key differences: (i) selective aggregation is applied in each round after a specified threshold, and (ii) the client update step involves training an unsupervised AD model. The threshold is not a sensitive parameter and is determined empirically. 

The process begins with the server initializing a global model. Let \(S\) denote the set of all clients. In each round, the server broadcasts the current global model to all clients.  Each client trains the model locally on its own dataset and sends the resulting model update to the server. In the initial rounds, before selective aggregation is activated, the server aggregates updates from all clients in \(S\). After a predefined warm-up period, the server selects a subset of clients \(S^t \subseteq S\) for aggregation by excluding clients suspected of containing anomalous samples. Finally, the server aggregates the subset \(S^t\subseteq S\) updates—weighted by the size of each client’s dataset—to update the global model. This process repeats iteratively until the end.

Figure~\ref{fig:fl-anomaly-moderation} shows the model updates during the iterative global model update. Initially, the distances between the global model and the updates from both benign and compromised clients are comparable, leading to insufficient refinement. However, due to the majority presence of benign clients, the aggregated updates predominantly reflect their contributions. Over successive iterations, models trained on compromised clients increasingly diverge from the global model, facilitating effective clustering-based client type estimation. This enables the exclusion of compromised clients from the training process. Ultimately, the global model is constructed by aggregating benign-client models.

\subsection{Unsupervised AD Based on Deep SVDD}
In this paper, we employ a well-known AD method, Deep Support Vector Data Description (Deep SVDD). Introduced by Ruff \textit{et al.}~\cite{pmlr-v80-ruff18a}, Deep SVDD represents the first fully deep learning-based one-class classification objective aimed at unsupervised AD.
While FLANDRE is adaptable to various AD models, Deep SVDD offers a computationally efficient and straightforward implementation in scenarios involving the removal of compromised clients, making it a favorable choice for NIDS.

\par Assume an input space \(X \subseteq \mathbb{R}^d\) and an output space \(F \subseteq \mathbb{R}^p\), and let \(\phi(\cdot; w): X \rightarrow F\) be a neural network with \(L \in \mathbb{N}\) hidden layers and a set of weights \(w = \{\bm{w}^1, \dots, \bm{w}^L\}\), where \(\bm{w}^\ell\) are the weights of layer \(\ell \in \{1, \dots, L\}\). Here, \(\phi(x; w) \in F\) represents the feature mapping of \(x \in X\) determined by the network \(\phi\) with parameters \(w\).

The goal of Deep SVDD is to jointly learn the network parameters \(w\) while minimizing the volume of a data-enclosing hypersphere in output space \(F\), characterized by a radius \(R > 0\) and a center \(c \in F\), which we assume to be predefined for now. Given some training data \(D = \{x_1, \dots, x_n\}\) on \(X\), we define the soft-boundary Deep SVDD objective as:
\begin{align}\label{eq:deepsvdd}
  \begin{split}
    \min _{R, w} R^2+\frac{1}{\nu n} \sum_{i=1}^n \max \left\lbrace0,\lVert\phi\left(\bm{x}_i; w\right)-c\rVert^2-R^2\right\rbrace \\+\frac{\lambda}{2} \sum_{\ell=1}^L\lVert\bm{w}^{\ell}\rVert_F^2.
  \end{split}
\end{align}

Minimizing \(R^2\) reduces the volume of the hypersphere. The second term is a penalty for points that, after being processed by the network, lie outside the sphere, i.e., if their distance to the center \(\|\phi(x_i; w) - c\|\) exceeds \(R\). The hyperparameter \(\nu \in \linterval{0}{1}\) moderates the trade-off between the volume of the sphere and boundary violations, allowing some points to be mapped outside the sphere.  The final term is a weight decay regularizer on the network parameters \(w\) with a hyperparameter \(\lambda > 0\), where \(\|\cdot\|_F\) denotes the Frobenius norm. Optimizing objective~\eqref{eq:deepsvdd} enables the network to learn parameters \(w\) such that data points are closely aligned to the center \(c\) of the hypersphere. To achieve this, the network must extract common factors of variation from the data.

\begin{figure}[t]
  \centering

  \includegraphics[width=0.7\columnwidth]{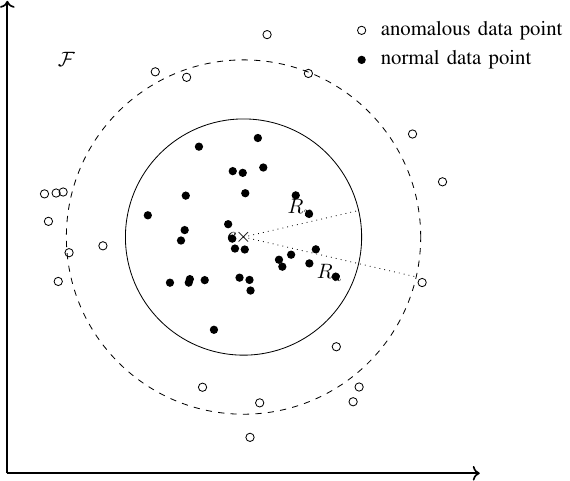}
  \caption{The conceptual framework wherein Deep SVDD establishes an optimal radius within the feature space, $\mathcal{F}$, effectively enabling the exclusion of anomalous data points.}
  \label{fig:deepsvdd}
\end{figure}

As illustrated in Figure~\ref{fig:deepsvdd}, normal examples of the data are closely mapped to the center \(c\), while anomalous examples are positioned further from the center or outside the hypersphere. This approach provides a compact description of the normal class and enforces the learning process through minimizing the size of the sphere.

\subsection{Model Training to Mitigate Effect of Contamination} 
\subsubsection{FL-aided unsupervised NIDS model training}
In this section, we describe our model training method based on the widely used FL algorithm, FedAvg. The primary goal is defined by the following distributed objective:
\begin{equation}
  \min_w \left\lbrace f(w) \coloneq \sum_{i \in  S} \frac{\lvert D_i \rvert}{\lvert D_S\rvert} L(D_i; w) \right\rbrace, 
\end{equation}
where \( S \) denotes the set of clients participating in the training process, \(\lvert D_i\rvert\) represents the size of client \( i \)'s dataset \( D_i \), \(\lvert D_S\rvert \coloneq\sum_{i\in S} \lvert D_i\rvert\),  and \(L( \cdot ; \cdot )\) is a user-specified loss function based on  \(D_i \) with model parameters \(w\). The global objective, \(f(w) \), acts as a collective goal for multiple clients, which FL seeks to minimize by optimizing all local objectives.

The FLANDRE algorithm is outlined in Algorithm~\ref{alg:fedavg}, where \( \lvert D_i\rvert \) represents the size of the training data for client \( i \), and the total size of the training data summed across clients in \(S^t\) is \(\lvert D_{S^t}\rvert \coloneq \sum_{i\in S^t} \lvert D_i \rvert \). During communication round \( t \), the server broadcasts the latest global model \( w^t \) to all the clients for local training. Each client proceeds to fine-tune the local objective function using its own data. Following local training, the parameter updates \( \Delta w_i^t \) computed by the clients who have successfully completed training within the allotted time frame are collected and aggregated by the server. This aggregation updates the global model to \( w^{t+1} \) as follows:

\begin{align}
  w^{t+1} = w^t + \frac{1}{|D_{S^t}|} \sum_{i \in S^t} |D_i| \Delta w_i^t,
\end{align}
where the update \( \Delta w \) is given by
\begin{align}
  \Delta w = -\eta \nabla L(D_i; w).
\end{align}
Achieving the desired accuracy in FL typically requires hundreds of rounds. Since our proposed method is unsupervised, the stopping condition should ideally be determined by whether the magnitude of the gradient falls below a certain threshold. In our experiments, we inspect the learning curves over a fixed number of communication rounds to analyze training dynamics. In this study, we also report oracle best-F1 values only to compare the achievable performance of different methods under the same evaluation protocol.

\subsubsection{Selective Aggregation via Model Clustering}
Our proposed method selectively excludes potentially compromised clients from model aggregation in order to mitigate their adverse impact on the global model. The key idea is that, because anomalous samples appear only at compromised clients, their locally updated models tend to deviate more from the global model than those of benign clients. FLANDRE exploits this deviation to identify and isolate compromised clients in a data-driven manner.

Selective aggregation is performed in two steps: (i) clustering of client models and (ii) selection of the cluster to be aggregated.

\noindent\textbf{Step 1: Clustering client models.}
As described in the system model, we assume that the participating clients consist of benign and compromised clients. Their training data distributions differ substantially; in particular, only the datasets of compromised clients contain anomalous samples. Consequently, after local training, the model parameters of compromised clients are expected to differ significantly from those of benign clients. To capture this discrepancy, we apply a clustering algorithm to the client models and partition them into two clusters, which are expected to roughly correspond to groups dominated by benign and compromised clients, respectively.

Various clustering algorithms can be employed in FLANDRE. Among them, the EM algorithm~\cite{reynolds2009gaussian} within the Gaussian mixture model (GMM) framework is one of the most established choices. It is widely used to obtain maximum-likelihood estimates and has demonstrated robust performance across a variety of practical applications~\cite{9201345, WANG2019154}. 
 
The EM algorithm is an iterative procedure that estimates model parameters to maximize the likelihood function $p(X \mid \lambda)$, where $\lambda$ denotes model parameters. Starting from an initial model $\lambda$, the algorithm iteratively updates the parameters to obtain a new estimate $\bar{\lambda}$ such that $p(X \mid \bar{\lambda}) \geq p(X \mid \lambda)$. The updated model $\bar{\lambda}$ then serves as the initial model for the next iteration, and this process is repeated until a predefined convergence criterion is satisfied.

In FLANDRE, we define distance $d_i$ of client $i \in S$ as follows:
\[
d_i \coloneq  \left\lVert w_i^t - w^{t-1} \right\rVert
\]
where $w_i^{t}$ is client $i$'s model and $w^{t-1}$ is the global model stored at the parameter server from the previous round. 
The scalar distance \(d_i\) is used as a one-dimensional input to the GMM, and we set the number of mixture components to $M = 2$. During each iteration of the EM algorithm, the following re-estimation formulas are applied; these guarantee a monotonic increase in the likelihood:\\
\textit{Mixture weights}
\[
  \bar{\phi}_k = \frac{1}{\lvert S\rvert} \sum_{i\in S} \Pr(k \mid d_i, \lambda).
\]
\textit{Means}
\[
  \overline{\mu}_k = \frac{\sum_{i\in S} \Pr(k \mid d_i, \lambda) d_i}{\sum_{i\in S} \Pr(k \mid d_i, \lambda)}.
\]
\textit{Variances (diagonal covariance)}
\[
  \bar{\sigma}_k^2 = \frac{\sum_{i\in S} \Pr(k \mid d_i, \lambda) d_i^2}{\sum_{i\in S} \Pr(k \mid d_i, \lambda)} - \bar{\mu}_k^2,
\]
where $\bar{\phi}_k$, $\bar{\mu}_k$ and $\bar{\sigma}_k$ denote the mixture weights and means, variances after re-estimation, respectively. The posterior probability $\Pr(k \mid d_i, \lambda)$ is computed as
\begin{align}
  \begin{split}
    \Pr(k \mid d_i, \lambda)
    & = \frac{\phi_k\, g(d_i \mid \mu_k, \Sigma_k)}{p(d_i\mid \lambda)} \\
    & = \frac{\phi_k\, g(d_i \mid \mu_k, \Sigma_k)}{\sum_{l=1}^M \phi_l\, g(d_i \mid \mu_l, \Sigma_l)},
  \end{split}
\end{align}
where $g(d_i \mid \mu_k, \Sigma_k)$ denotes the Gaussian probability density function parameterized by the mean $\mu_k$ and covariance $\Sigma_k$ for the $k$-th component, and $p(d_i\mid\lambda)$ is the likelihood of $d_i$ under the GMM with parameters $\lambda$.

After convergence, each client model is assigned to one of the two clusters according to the maximum a posteriori rule
\[
  k^\ast(i) = \arg\max_{k \in \{0,1\}} \Pr(k \mid d_i, \lambda),
\]
and we define the corresponding client index sets as
\[
  S_k = \{\, i\in S \mid k = k^\ast(i) \,\}, \quad k \in \{0,1\}.
\]
These two sets $S_0$ and $S_1$ are then used in the subsequent cluster selection step.

\vspace{2mm}
\noindent\textbf{Step 2: Selecting the cluster for aggregation.}
Next, we evaluate each cluster and decide which one should be excluded from aggregation. To this end, we leverage the distance between client models and the global model. Over successive training rounds, models trained on compromised clients typically diverge more from the global model than those trained on benign clients. Let $w_i^{t}$ denote the locally updated model at client $i$ in round $t$ and $w^{t-1}$ denote the global model from the previous round. In general, the Euclidean distance between $w_i^{t}$ and $w^{t-1}$ for compromised clients is larger than that for benign clients.

For each cluster $S_k$, we therefore compute the average model distance
\begin{align}\label{eq:dist}
  d(S_k) = \frac{1}{\lvert S_k\rvert}\sum_{i\in S_k} \left\lVert w_i^t - w^{t-1} \right\rVert= \frac{1}{\lvert S_k\rvert}\sum_{i\in S_k} d_i,
\end{align}
where $w_i^{t}$ is client $i$'s model and $w^{t-1}$ is the global model stored at the parameter server from the previous round. We regard the cluster with the larger average distance as containing a higher proportion of compromised clients and exclude it from model aggregation. The remaining cluster, whose distance $d(S_k)$ is smaller, is treated as the benign-dominated cluster and is used to update the global model $w^t$.

\noindent\textbf{Option: Hypersphere radius of Deep SVDD.} When Deep SVDD is used as the underlying anomaly detector, the hypersphere radius $R_i$ of the soft-boundary can be employed to replace $d_i$ in both step 1 and 2. Specifically, instead of evaluating the distance to the global model, we compute the average radius within each cluster, $\sum_{i\in S_k} R_i/\lvert S_k\rvert$, and interpret the cluster with the larger average radius as containing more compromised clients, which is then excluded from aggregation. A larger $R_i$ indicates that the training data of client $i$ are more widely spread in the embedding space of Deep SVDD, which is often a consequence of the influence of anomalous samples. An important advantage of this FLANDRE-R criterion is that it does not require collecting the entire model from each client, thereby reducing communication overhead during training. However, as shown later in the experimental evaluation, the FLANDRE-R criterion can be unstable and may fail to discriminate compromised clients in some tasks.

We designate the method that employs the Euclidean distance of model parameters simply as \textit{FLANDRE}, and the variant that utilizes the Deep SVDD parameter $R$ for selective aggregation as \textit{FLANDRE-R}.

\section{Performance Evaluation}\label{sec:evaluation}
\subsection{Evaluation Setup}

\subsubsection{Scenario}

We evaluate FLANDRE in a simulated FL environment that emulates an IoT deployment of NIDS using publicly available NIDS benchmark datasets. All clients are simulated on a single server; network latency and inter-client communication overhead are not in scope of this study and are therefore ignored. The total number of clients is fixed to \(N = 100\), and we assume full participation in every federated round, i.e., all clients perform local training and send updates to the server. For FLANDRE, however, only the client updates selected by the selective aggregation mechanism are used to update the global model.

In the performance evaluation, we adopt the same assumptions as in the system model and consider two types of clients: benign clients, which store only normal traffic, and compromised clients, whose local datasets contain both normal and anomalous traffic. The parameter \(r\) denotes the \emph{contamination ratio} in the dataset. The parameter \(\gamma\) denotes the \emph{ratio of anomalous samples in the local dataset of a compromised client}. Thus, each compromised client holds a fixed-size local dataset in which a proportion \(\gamma\) of the samples are anomalous and the rest are normal, whereas each benign client holds only normal samples.

Table~\ref{table:hyper_param} summarizes the main hyperparameters used in the federated setting. Most hyperparameters are tuned using the centralized Deep SVDD model on the ToN\_IoT dataset and then reused across all experiments. In the federated setup, the learning rate used for centralized training was found to lead to slow convergence; therefore, the learning rate \(\alpha\) is increased by a factor of ten for all federated experiments, while the other hyperparameters remain unchanged.

To construct the client local datasets, we first separate the training split into anomalous and benign samples. Let \(\mathcal{A}\) and \(\mathcal{B}\) denote the sets of anomalous and benign training samples, respectively, where the overall contamination ratio is approximately \(r = |\mathcal{A}|/(|\mathcal{A}|+|\mathcal{B}|)\). Given the target anomalous-sample ratio \(\gamma\) in compromised clients, we construct a compromised-client data pool by combining all anomalous samples with
\[
\left\lfloor |\mathcal{A}| \frac{1-\gamma}{\gamma} \right\rfloor
\]
benign samples drawn from \(\mathcal{B}\). The remaining benign samples form the benign-client data pool. This construction makes the anomalous-sample ratio in the compromised-client pool approximately \(\gamma\). Since the size of the compromised-client pool is approximately \(|\mathcal{A}|/\gamma\), the fraction of compromised clients is approximately \(r/\gamma\). We then partition the benign-client pool and the compromised-client pool separately into disjoint subsets, whose numbers are proportional to the sizes of the two pools, and assign each subset to one client. Consequently, benign clients contain only benign samples, whereas compromised clients contain a mixture of anomalous and benign samples with an anomalous-sample ratio of approximately \(\gamma\).

\begin{table}[t]
  \centering
  \begin{tabular}{l|c}% chktex-file 44
    \toprule
    number of clients \(N\)                                   & \(100\)                \\
    contamination ratio \(r\)                     & \(0.1 \mbox{--} 0.40\)   \\
    anomalous-sample ratio in compromised clients \(\gamma\)  & \(20\mbox{--}80\%\)    \\
    learning rate (federated) \(\alpha\)                      & \(3.90\times 10^{-4}\) \\
    weight decay \(\beta\)                                    & \(7.64\times 10^{-7}\) \\
    learning-rate milestone                                   & \(50\)                 \\
    Deep SVDD hyperparameter \(\nu\)                          & \(0.975\)              \\
    number of federated rounds \(K\)                          & \(100\)                \\
    \bottomrule
  \end{tabular}
  \caption{Experimental configuration in the FL setting. For the centralized baselines, we use the same hyperparameters except for the learning rate, which is set to \(\alpha = 3.90 \times 10^{-5}\). The ranges of \(r\) and \(\gamma\) are varied in separate experiments.}
  \label{table:hyper_param}
\end{table}

\subsubsection{Datasets}

We evaluate FLANDRE on three recent NIDS benchmarks: ToN\_IoT~\cite{toniot}, IDS2018~\cite{sharafaldinGeneratingNewIntrusion2018}, and NF-UQ-NIDS~\cite{sarhanStandardFeatureSet2022}. In all cases, we use the flow-based representations and apply a unified preprocessing and partitioning procedure.

\textbf{ToN\_IoT} is a heterogeneous dataset introduced in 2019 that contains IoT/industrial Internet of Things (IIoT) telemetry, host logs, and network traffic. We use the network flow subset provided in the \textit{Train\_Test\_datasets} package, which is specifically designed for ML-based NIDS evaluation. This subset contains approximately \(50\mathrm{k}\) normal and \(161\mathrm{k}\) anomalous flows generated from a variety of attacks, including scanning, denial-of-service (DoS)/distributed denial-of-service (DDoS), ransomware, and backdoor attacks, executed in a large-scale cyber range environment. The original flow representation comprises 44 features extracted with Bro-IDS (now Zeek), including source and destination Internet Protocol (IP) addresses and ports, transport and application protocols, payload sizes, connection state, flow duration, packet counts, and statistics related to Domain Name System (DNS), Secure Sockets Layer (SSL), and Hypertext Transfer Protocol (HTTP) activity.

\textbf{IDS2018} is a collaborative dataset from the Communications Security Establishment and the Canadian Institute for Cybersecurity. It captures realistic benign and attack traffic over ten days and includes modern attack scenarios such as DDoS, DoS, brute-force, cross-site scripting (XSS), Structured Query Language (SQL) injection, botnet, and infiltration attacks. In our experiments, we use the network flows collected on March 23, 2018, which comprise about \(331\mathrm{k}\) flows, of which approximately \(72\%\) are normal. The flow features include duration, forward and backward packet counts, packet-size statistics, inter-arrival time statistics, header lengths, flow rates, and bulk transfer metrics.

\textbf{NF-UQ-NIDS} is a unified flow dataset constructed from UNSW-NB15, ToN\_IoT, BoT-IoT, and IDS2018. By harmonizing the feature sets of these benchmarks, it provides a comprehensive NIDS dataset covering diverse network configurations and attack contexts. An additional label indicates the original source dataset of each flow, enabling cross-dataset analysis. Attack categories are consolidated into broader classes such as DoS, DDoS, brute force, and injection. The features include Internet Protocol version 4 (IPv4) addresses and ports, protocol identifiers, byte and packet counts in each direction, flow duration, cumulative Transmission Control Protocol (TCP) flags, per-direction stream durations, minimum and maximum packet sizes, flow-level throughput statistics, and indicators for TCP, Internet Control Message Protocol (ICMP), DNS, and File Transfer Protocol (FTP) activity.

For all three datasets, we remove explicit identifiers such as IP addresses and timestamps before training. For features with a wide dynamic range (e.g., packet lengths), we apply a logarithmic transformation to reduce skewness. The resulting feature vectors are used as inputs to the models. 

\subsubsection{Implementation Details}

All experiments are implemented in Python. We use the Flower framework~\cite{beutel2020flower} to simulate the FL process, including client-side training and server-side aggregation. The simulations are executed on Amazon Elastic Compute Cloud (EC2) c5.4xlarge instances.

\subsection{Baselines}
To assess the effectiveness of FLANDRE, we compare it with both centralized and federated baselines that represent standard practice and SoTA AD approaches on NIDS dataset. Unless otherwise noted, all methods use the same backbone network as our proposed method, i.e., a feed-forward neural network with layer dimensions \(64 \rightarrow 32 \rightarrow 64\), so that performance differences primarily reflect the learning paradigm (centralized vs.\ federated) and the treatment of contaminated data rather than the architectural capacity of the models.

\textbf{Ideal Deep SVDD (no contamination):}
As a reference on achievable performance, we train a standard soft-boundary Deep SVDD model on a contamination-free training set composed only of normal samples. This idealized centralized setting assumes that no anomalous flows are present during training. Although difficult to realize in practice, it serves as a reference to quantify the degradation caused by contamination and by federated constraints. 

\textbf{Deep SVDD (Centralized):}
To evaluate the impact of contamination in a centralized setting that reflects the federated scenario, we construct a centralized training set by aggregating all client-local datasets described above. The centralized dataset contains anomalous samples in the contamination ratio \(r\). We then train a soft-boundary Deep SVDD model on this contaminated centralized dataset in a fully unsupervised manner. 

\textbf{LOE-S (Centralized):}
We also compare against the SoTA ML-based AD method proposed by Qiu et al.~\cite{pmlr-v162-qiu22b}, denoted as \textit{LOE-S (Centralized)}. For a fair comparison, LOE-S is trained on the same contaminated centralized dataset obtained by aggregating all client-local datasets, which again contains anomalous samples in the contamination ratio \(r\). LOE-S generates synthetic labeled anomalies and optimizes a joint loss that leverages both normal data and synthetically generated anomalous samples, and has been shown to be highly effective for AD under contamination. In our experiments, we adapt LOE-S to the same model architecture as used for Deep SVDD, so that all centralized methods share comparable model capacity.

\textbf{Federated Deep SVDD:}
To evaluate the effectiveness of the selective aggregation mechanism in FLANDRE, we also consider a baseline obtained by removing selective aggregation from FLANDRE and applying a standard FL framework to Deep SVDD. Similar to FLANDRE, this \textit{Federated Deep SVDD} baseline leverages federated training to mitigate the impact of anomalous sample contamination, but it does not perform any selective aggregation or client filtering. This setting allows us to isolate the additional benefit provided by FLANDRE’s selective aggregation mechanism. Furthermore, by comparing Federated Deep SVDD with the centralized Deep SVDD baseline, we can assess the advantage of employing FL itself under contamination.

\textbf{Proposed methods: FLANDRE and FLANDRE-R (optional).}
Our proposed method FLANDRE is built on the key idea of exploiting FL to reduce the impact of anomalous sample contamination. By aggregating models across many predominantly benign clients, FL naturally downweights the influence of a small number of compromised clients whose local data contain anomalies. On top of this effect, FLANDRE further introduces a selective aggregation mechanism based on the Euclidean distance between local models and the global model, explicitly suppressing updates from clients that deviate significantly and are likely to be compromised. This distance-based variant is reported simply as \textit{FLANDRE}. In addition, we consider an optional variant that uses the Deep SVDD hypersphere radius \(R\) instead of model parameters for client selection, denoted as \textit{FLANDRE-R (optional)}. Both FLANDRE and FLANDRE-R share the same backbone architecture and hyperparameters as the Deep SVDD-based baselines, allowing a fair comparison across all methods.

\subsection{Evaluation Results}
\begin{table*}[t]
  \centering
  \small
  \setlength{\tabcolsep}{7pt}
  \begin{tabular}{lccc}
    \toprule
    Method                           & ToN\_IoT                     & NF-UQ-NIDS                        & IDS2018                      \\
    \midrule
    \multicolumn{4}{l}{\textbf{Baselines}}                                             \\
    \quad Ideal Deep SVDD (no contamination)   & \(0.973 \pm 0.003\)          & \(0.831 \pm 0.013\)          & \(0.876 \pm 0.071\)          \\
    \quad Deep SVDD (Centralized)      & \(0.925 \pm 0.017\)          & \(0.797 \pm 0.035\)          & \(0.590 \pm 0.061\)          \\
    \quad                            & {\scriptsize\((-0.048)\)}    & {\scriptsize\((-0.034)\)}    & {\scriptsize\((-0.286)\)}    \\
    \quad LOE-S (Centralized)           & \(0.916 \pm 0.019\)          & \(0.742 \pm 0.034\)          & \(0.586 \pm 0.057\)          \\
    \quad                            & {\scriptsize\((-0.057)\)}    & {\scriptsize\((-0.089)\)}    & {\scriptsize\((-0.290)\)}    \\

    \quad Federated Deep SVDD
                                     & \(0.948 \pm 0.009\)          & \(0.779 \pm 0.025\)          & \(0.630 \pm 0.060\)          \\
     \quad                            & {\scriptsize\((-0.025)\)}    & {\scriptsize\((-0.052)\)}    & {\scriptsize\((-0.246)\)}    \\

                                         \midrule
    \multicolumn{4}{l}{\textbf{Proposed methods}}                                                                 \\
    
    \quad FLANDRE
                                     & \(\mathbf{0.969 \pm 0.002}\) & \(\mathbf{0.838 \pm 0.010}\) & \(\mathbf{0.823 \pm 0.038}\) \\
    \quad                            & \(\mathbf{\scriptstyle(-0.004)}\)
                                                                     & \(\mathbf{\scriptstyle(+0.007)}\)
                                                                                                         & \(\mathbf{\scriptstyle(-0.053)}\) \\
    \quad FLANDRE-R (optional)
                                     & \(0.968 \pm 0.003\)          & \(0.801 \pm 0.024\)          & \(0.790 \pm 0.059\)          \\
    \quad                            & {\scriptsize\((-0.005)\)}    & {\scriptsize\((-0.030)\)}    & {\scriptsize\((-0.086)\)}    \\
    \bottomrule
  \end{tabular}
  \caption{F1 scores on three datasets. ``Ideal (no contamination)'' denotes centralized training without contaminated samples. Numbers in parentheses show the difference from this ideal setting (method minus ideal). We set the contamination ratio \(r\) of the training set to \(0.1\) and the anomalous-sample ratio in compromised clients to \(\gamma=0.5\).}
  \label{table:dataset_diff}
\end{table*}

\begin{figure*}[t]
  \centering
  \includegraphics[width=\linewidth]{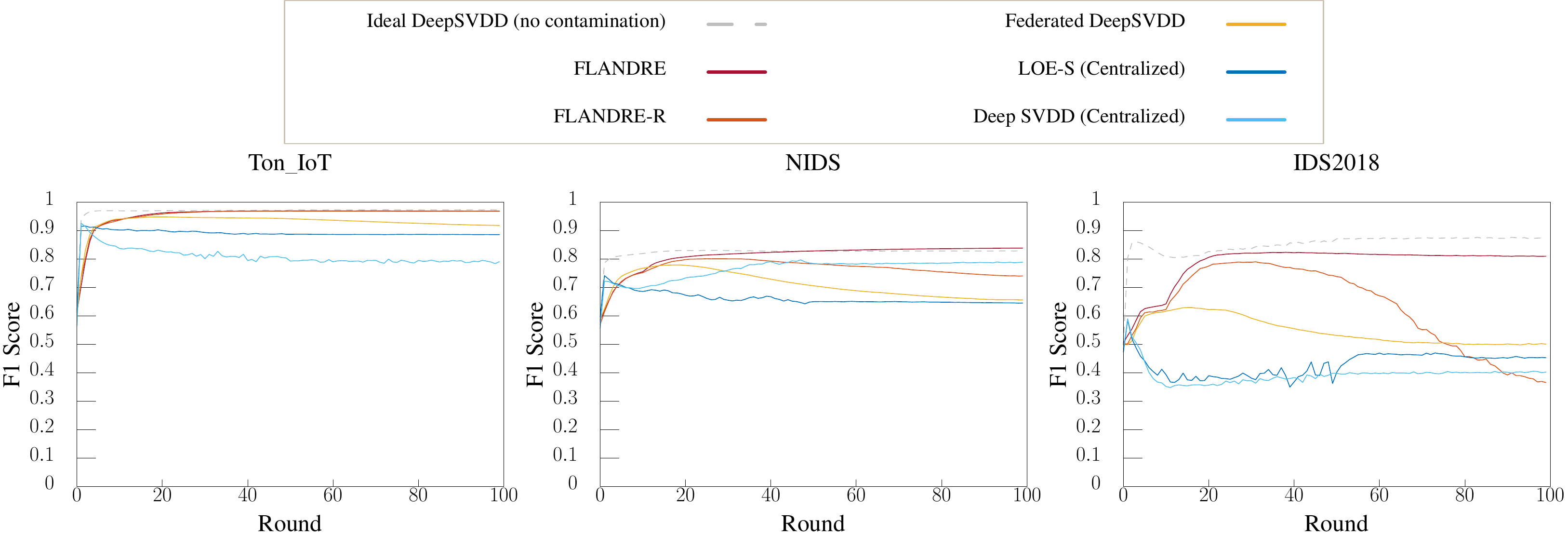}
  \caption{Learning curves (F1 score vs.\ communication rounds) on the three datasets under \(r = 0.1\) and \(\gamma = 0.5\).}
  \label{fig:curve}
\end{figure*}

\begin{figure*}[t]
  \centering
  \includegraphics[width=\linewidth]{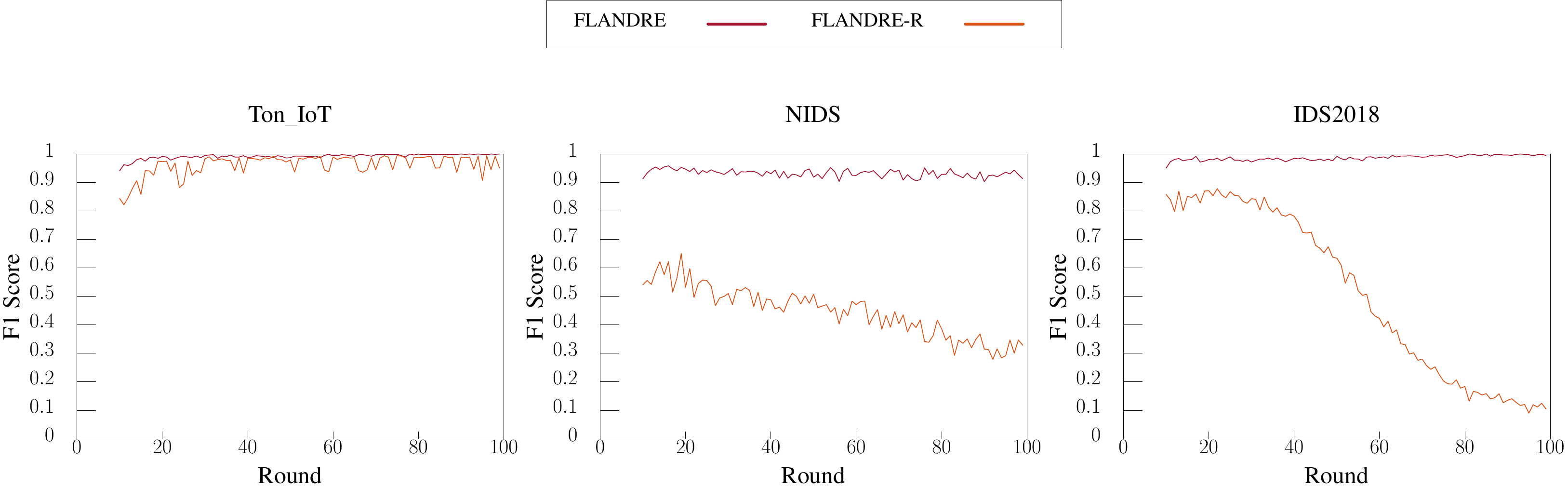}
  \caption{F1 score of selective aggregation for FLANDRE and FLANDRE-R under \(r = 0.1\) and \(\gamma = 0.5\).}
  \label{fig:r}
\end{figure*}

\subsubsection*{Performance across datasets}
Table~\ref{table:dataset_diff} reports the F1 scores on ToN\_IoT, NF-UQ-NIDS, and IDS2018 with the contamination ratio fixed to \(r = 0.1\) and the anomalous-sample ratio in compromised clients fixed to \(\gamma = 0.5\), i.e., \(10\%\) of the aggregated training samples are anomalous and half of a compromised client's local samples are anomalous. For each method and dataset, we report the mean and standard deviation of the \emph{best} F1 score over 20 runs, where the best F1 is defined as the maximum test F1 along the learning curve for that run. Although this oracle stopping point is not available in practice, it is applied uniformly to all methods and thus provides a fair comparison of their achievable performance. In addition, Fig.~\ref{fig:curve} shows that FLANDRE exhibits a stable learning curve, whereas some baseline methods show peaky trajectories and experience sharp performance degradation after a certain number of rounds. Therefore, this evaluation protocol does not favor the proposed method. The values in parentheses indicate the difference from \textit{Ideal Deep SVDD (no contamination)}.

From Table~\ref{table:dataset_diff}, FLANDRE achieves the highest F1 scores on all three datasets and remains very close to the ideal contamination-free reference. On ToN\_IoT, FLANDRE attains an F1 of \(0.969\), only \(0.004\) below \textit{Ideal Deep SVDD}, and clearly surpasses Deep SVDD (Centralized), LOE-S (Centralized), and Federated Deep SVDD. A similar gap in favor of FLANDRE is observed on NF-UQ-NIDS and IDS2018, indicating that combining FL with distance-based selective aggregation is highly effective in mitigating client-level contamination. The optional variant FLANDRE-R also improves significantly over the baselines but is consistently slightly worse than FLANDRE, especially on NF-UQ-NIDS and IDS2018, suggesting that the \(R\)-based selection criterion is less robust.

Figure~\ref{fig:curve} shows the corresponding learning curves (test F1 over communication rounds). While the table focuses on the best F1 values, the curves reveal the training dynamics. FLANDRE exhibits smooth and stable trajectories and quickly reaches a broad high-performance plateau, implying that reasonable early-stopping rules could achieve near-optimal performance without precise tuning. In contrast, the baselines, particularly centralized Deep SVDD, often show sharp peaks followed by degradation, making it difficult to exploit their best F1 in a fully unsupervised setting. FLANDRE-R generally follows FLANDRE but exhibits more fluctuation and overfitting on NF-UQ-NIDS and IDS2018, in line with its lower best F1 scores.

Figure~\ref{fig:r} further analyzes the selective aggregation mechanism by reporting the F1 scores for classifying clients into benign and compromised groups. On ToN\_IoT, both FLANDRE and FLANDRE-R almost perfectly separate compromised clients from benign ones. On NF-UQ-NIDS and IDS2018, the distance-based FLANDRE maintains high client-selection accuracy, whereas the \(R\)-based FLANDRE-R is clearly less reliable. This strong client-selection capability of FLANDRE explains why it can construct robust global models and achieve the high detection performance summarized in Table~\ref{table:dataset_diff}.

\begin{figure}[t]
  \centering

  \includegraphics[width=\columnwidth]{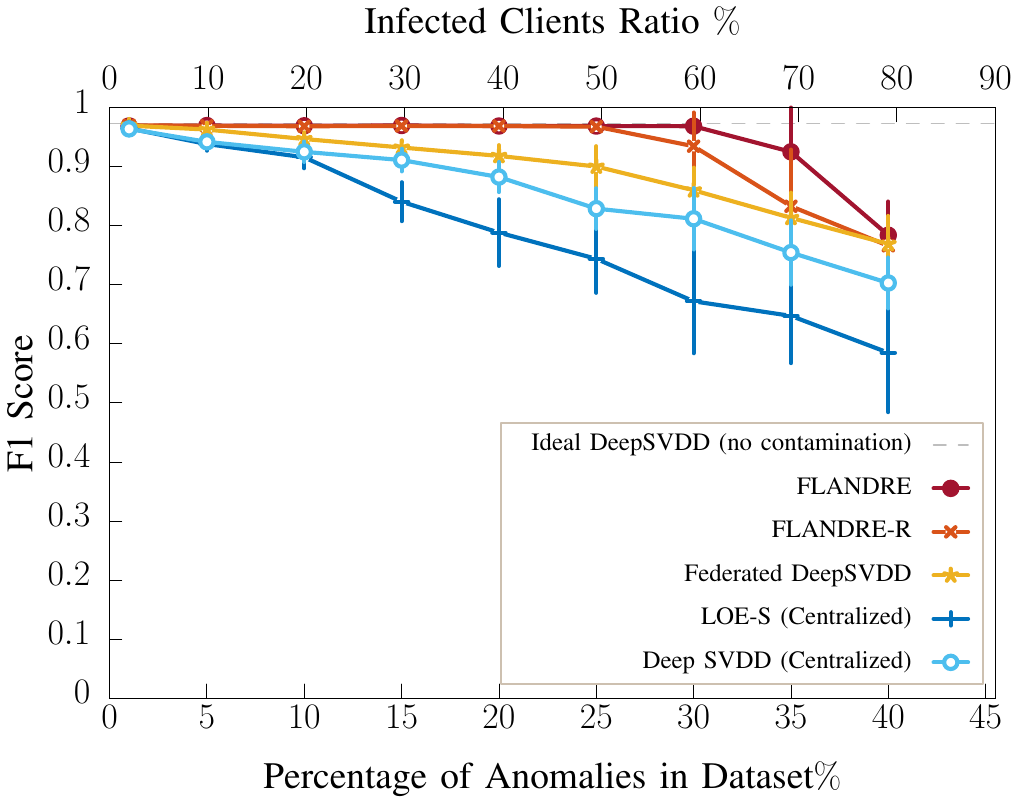}
  \caption{Performance analysis across different contamination ratios on the ToN\_IoT dataset. We set the  anomalous-sample ratio in compromised clients \(\gamma\) as \(\gamma=0.5\). }
  \label{fig:contami}
\end{figure}

\subsubsection*{Across Different Contamination Ratios}
Figure~\ref{fig:contami} shows the F1 scores for various contamination ratios.
In these experiments, we investigated the impact of varying contamination ratios within the dataset on model performance. We selected the ToN\_IoT dataset due to its widespread recognition and prominence in network AD research. The anomalous-sample ratio in compromised clients was set to $\gamma = 0.5$, thereby yielding a compromised-client ratio that is twice the contamination rate $r$. We report the mean and standard deviation of the F1 scores over 20 independent trials, each using different model initializations and distinct anomalous samples for contamination.
\par Our proposed method demonstrates robust performance under high contamination ratios, particularly when the compromised-client ratio remains below \(50\%\). Federating the basic Deep SVDD model yields some performance improvements, but these gains remain relatively limited. In contrast, our proposed method demonstrates superior results, with both the FLANDRE and FLANDRE-R approaches outperforming the basic Deep SVDD by approximately 0.14 in terms of F1 score. Furthermore, the performance degradation from the ideal state is constrained to within 0.005, indicating the robustness of our approach. Although FLANDRE demonstrates superior robustness against anomalous data contamination compared to the \( R \)-based method when the contamination ratio \( r \) exceeds 35\%, their performance remains largely comparable within practical contamination levels. The performance of LOE-S consistently falls below that of basic Deep SVDD\@. This can be attributed to the fact that basic Deep SVDD exhibits a sharp, peaky learning curve.

\begin{figure}[t]
  \centering

  \includegraphics[width=\columnwidth]{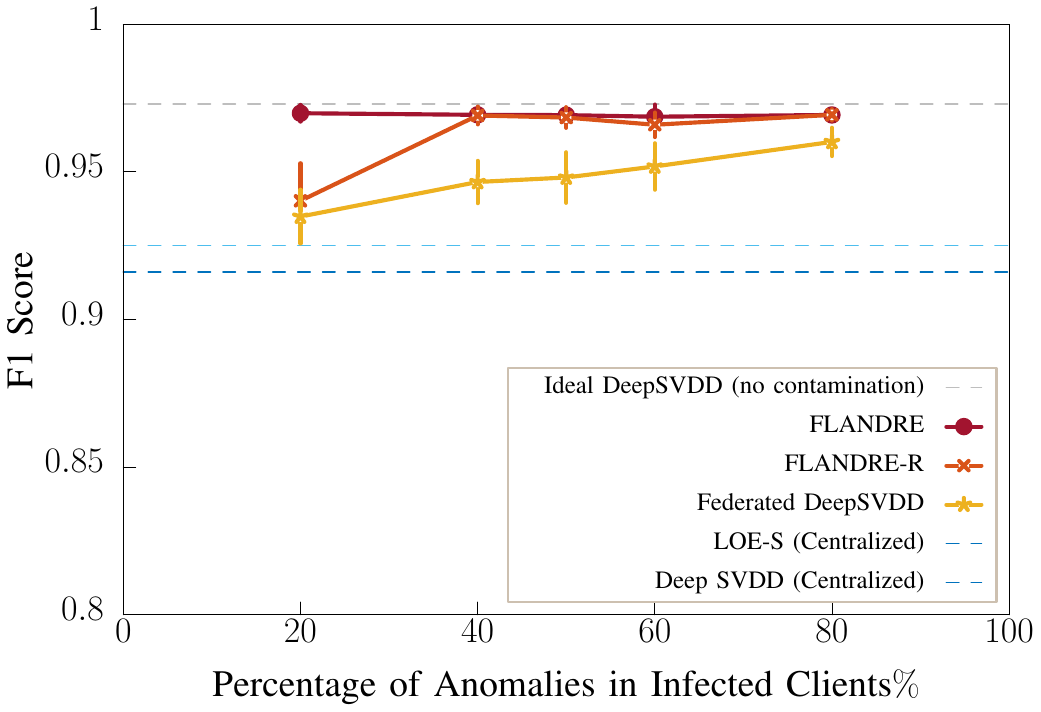}
  \caption{Performance analysis across different anomalous-sample ratios in compromised clients on ToN\_IoT dataset. There is an inverse relationship between the anomalous-sample ratio within compromised clients and the compromised-client ratio. }
  \label{fig:duty}
\end{figure}

\subsubsection*{Across Different Anomalous-sample Ratios in Compromised Clients}
Figure~\ref{fig:duty} shows the variation in performance when the proportion of anomalous data held by compromised clients changes, even though the overall proportion of anomalous data within the network remains constant.
The ToN\_IoT dataset was chosen due to its established recognition and significance within the domain of network AD research. For these experiments, the contamination ratio was configured to $r = 0.1$, while the compromised-client ratio was determined based on the anomalous-sample ratio in the compromised clients, denoted as $\gamma$. We report the mean and standard deviation of the F1 scores over 20 independent trials, each using different model initializations and distinct anomalous samples for contamination.
\par FLANDRE-R  exhibits a slight performance degradation when the proportion of anomalous data within compromised clients is low. This degradation is likely attributed to the increase in the proportion of compromised clients as the contamination ratio $r$ remains fixed. Because all clients are assigned the same number of local samples and benign clients contain no anomalous samples, the fraction of compromised clients is determined by the overall contamination ratio \(r\) and the anomalous-sample ratio \(\gamma\) within compromised clients as \(r/\gamma\). On the other hand, FLANDRE appears largely unaffected by the proportion of anomalous data within compromised clients, demonstrating robustness against such variations.

\subsubsection*{Training Cost Evaluation}
FLANDRE incurs a training cost comparable to that of FedAvg. 
The training cost can be broadly decomposed into two primary components: 
communication cost and computational cost.

Regarding the communication cost, the number of model parameters depends on the input dimensionality. 
In our experiments, the model contains approximately 5--7K trainable parameters. 
When stored in float32 format without compression, a single model update requires approximately 20--30 KB. 
Since each participating client downloads the global model and uploads its local model in each communication round, 
the communication cost per round is approximately 40--60 KB per client. 
In our experimental setting, 100 communication rounds are performed, resulting in a total communication volume of approximately 4.0--6.0 MB per client over the entire training process. 
Equivalently, with 100 clients, the server-side aggregate communication volume is approximately 4--6 MB per round and 400--600 MB over the full training process. 
Furthermore, a variety of model compression techniques for reducing communication overhead have been extensively studied; applying such techniques can further reduce this cost.

Regarding the computational cost, FLANDRE is lightweight because it employs a small neural network and operates on tabular traffic features. 
Therefore, graphics processing unit (GPU) acceleration or specialized hardware is not required. 
In our experiments, training was conducted on an Amazon EC2 c5.4xlarge instance with 16 virtual central processing units (vCPUs). 
Each client process used a single vCPU, and up to 16 clients were processed in parallel. 
Under this setting, the total wall-clock training time for FLANDRE with 100 clients was approximately 15 minutes. 
For reference, centralized training of LOE-S required approximately 7 minutes under the same hardware configuration. 
Compared with standard FedAvg, the additional server-side computation in FLANDRE consists only of computing client-level model distances and applying GMM-based clustering over the clients. 
Since this operation scales with the number of clients rather than the number of traffic samples, its overhead is negligible compared with local Deep SVDD training.

\section{Conclusion}\label{sec:conclusion}
We propose FLANDRE, a contamination-robust unsupervised training framework for NIDS, tailored for real-world environments where some clients may be compromised. By leveraging FL's inherent tendency to underrepresent minority data---often considered a drawback---we mitigate the impact of data contamination caused by a limited number of compromised clients. Moreover, to avoid performance degradation, we exclude clients suspected of containing anomalous data from model aggregation. Extensive evaluations on publicly available NIDS datasets demonstrate that FLANDRE consistently outperforms baseline methods and remains robust even under high contamination ratios. As future work, we will provide a theoretical analysis of how much FL can mitigate contamination in AD beyond the qualitative insights and empirical validation presented in this paper. We will also investigate client heterogeneity, particularly non-independent and identically distributed (non-IID) normal traffic across clients, and develop methods to alleviate the adverse effects of non-IID data.

\bibliographystyle{IEEEtran.bst}
\bibliography{FL-AD.bib}

\end{document}